\documentclass[12pt]{iopart}
\usepackage{iopams}
\usepackage{setstack}
\usepackage{graphicx}

\newcommand{\bfe}{\bi{e}}
\newcommand{\eee}{\bi{e}}
\newcommand{\JJ}{\tilde{J}}
\newcommand{\JJJ}{\bi{J}}
\newcommand{\KKK}{\bi{K}}
\newcommand{\bfp}{\bi{p}}
\newcommand{\QQ}{\tilde{Q}}
\newcommand{\QQQ}{\bi{Q}}
\newcommand{\vvv}{\bi{v}}
\newcommand{\bfv}{\bi{v}}
\newcommand{\xxx}{\bi{x}}
\newcommand{\ppphi}{\bphi}

\begin{document}

\title{Gauge Invariance from a Graphical Self-Consistency Criterion}

\author{W.M. Stuckey$^1$, T.J. McDevitt$^2$ and M. Silberstein$^3$}

\address{$^1$ Department of Physics \\ Elizabethtown College \\ Elizabethtown, PA  17022}
\address{$^2$ Department of Mathematical Sciences \\ Elizabethtown College \\ Elizabethtown, PA  17022}
\address{$^3$ Department of Philosophy \\ Elizabethtown College \\ Elizabethtown, PA  17022}
\ead{\mailto{stuckeym@etown.edu}, \mailto{mcdevittt@etown.edu}, \mailto{silbermd@etown.edu}}

\begin{abstract}
We propose that quantum physics is the continuous approximation of a more fundamental, discrete graph theory (theory X).  Accordingly, the Euclidean transition amplitude $Z$ provides a partition function for geometries over the graph, which is characterized topologically by the difference matrix and source vector of the discrete graphical action. The difference matrix and source vector of theory X are related via a graphical self-consistency criterion (SCC) based on
the boundary of a boundary principle on a graph ($\partial_1\cdot \partial_2 = 0$). In this approach, the SCC ensures the source vector is divergence-free and resides in the row space of the difference matrix. Accordingly, the difference matrix will necessarily have a nontrivial eigenvector with eigenvalue zero, so the graphical SCC is the origin of gauge invariance.  Factors of infinity associated with gauge groups of infinite volume are excluded in our approach, since $Z$ is restricted to the row space of the difference matrix and source vector. Using this formalism, we obtain the two-source Euclidean transition amplitude over a ($1+1$)-dimensional graph with $N$ vertices fundamental to the
scalar Gaussian theory.
\end{abstract}
\noindent{\it Keywords \/}:graph theory, path integral, gauge invariance, transition amplitude
\pacs{03.65.Ca; 03.65.Ta;03.65.Ud; 11.15.-q}


\thispagestyle{empty}

\section{Introduction}
\label{section1}

Those who emphasize the incompleteness of quantum field theory (QFT)
over its successes often focus on the many ad hoc and, for some,
troubling ``fixes'' involved in the practice of QFT \footnote{We are
focusing on the ``textbook variant of QFT." Fraser, D.: Quantum
Field Theory: Underdetermination, Inconsistency, and Idealization.
Philosophy of Science 74, 536-565 (October 2009).}. For example,
since QFT is independent of overall factors in the transition
amplitude, such factors are simply ``thrown away'' even when these
factors are infinity as is the case when the volume of the gauge
symmetry group in Faddeev-Popov gauge fixing is infinite\cite{zee}.
In a petition to philosophers of science, Glashow
stated\cite{glashow}, ``in a sense it really is a time for people
like you, philosophers, to contemplate not where we're going,
because we don't really know and you hear all kinds of strange
views, but where we are. And maybe the time has come for you to tell
us where we are.'' Rovelli went further stating\cite{rovelli}, ``As
a physicist involved in this effort, I wish that the philosophers
who are interested in the scientific description of the world would
not confine themselves to commenting and polishing the present
fragmentary physical theories, but would take the risk of trying to
look ahead.''

Of course, ignoring factors of infinity in the transition amplitude
$Z$ per Faddeev-Popov gauge fixing is easily understood in terms of
(infinitely) over counting gauge degrees of freedom in the classical
field being quantized\cite{kaku}, so there is no problem in that
respect. We believe the real issue is the fact that QFT involves the
quantization of a classical field\cite{wallace} when one would
rather expect QFT to originate independently of classical field
theory, the former typically understood as fundamental to the
latter. Herein we accept Glashow and Rovelli's challenges and
respond, not philosophically, but mathematically, and propose a new,
fundamental origin for QFT. Specifically, we follow the possibility
articulated by Wallace\cite{wallace} that (p 45), ``QFTs as a whole
are to be regarded only as approximate descriptions of some
as-yet-unknown deeper theory,'' which he calls ``theory X," and we
propose a new discrete path integral formalism over graphs for
``theory X'' underlying QFT. Accordingly, sources $\JJJ$ , space and
time are self-consistently co-constructed per a graphical
self-consistency criterion (SCC) based on the boundary of a boundary
principle\cite{misner} on the graph ($\partial_1\cdot
\partial_2 = 0$). [In a graphical representation of QFT, part of $\JJJ$
represents field disturbances emanating from a source location
(Source) and the other part represents field disturbances incident
on a source location (sink).] We call this amalgam
"spacetimematter." The SCC constrains the difference matrix and
source vector in Z, which then provides the probability for finding
a particular source-to-source relationship in a quantum experiment,
i.e., experiments which probe individual source-to-source relations
(modeled by individual graphical links) as evidenced by discrete
outcomes, such as detector clicks. Since, in QFT, all elements of an
experiment, e.g., beam splitters, mirrors, and detectors, are
represented by interacting sources, we confine ourselves to the
discussion of such controlled circumstances where the empirical
results evidence individual graphical links. [Hereafter, all
reference to ``experiments'' will be to ``quantum experiments.''] In
this approach, the SCC ensures the source vector is divergence-free
and resides in the row space of the difference matrix, so the
difference matrix will necessarily have a nontrivial eigenvector
with eigenvalue zero, a formal characterization of gauge invariance.
Thus, our proposed approach to theory X provides an underlying
origin for QFT, accounts naturally for gauge invariance, i.e., via a
graphical self-consistency criterion, and excludes factors of
infinity associated with gauge groups of infinite volume, since the
transition amplitude $Z$ is restricted to the row space of the
difference matrix and source vector.

While the formalism we propose for theory X is only suggestive, the
computations are daunting, as will be evident when we present the
rather involved graphical analysis underlying the Gaussian
two-source amplitude which, by contrast, is a trivial problem in its
QFT continuum approximation. However, this approach is not intended
to replace or augment QFT computations. Rather, our proposed theory
X is fundamental to QFT and constitutes a new program for physics,
much as quantum physics relates to classical physics. Therefore, the
motivation for our theory X is, at this point, conceptual and while
there are many conceptual arguments to be made for our
approach\cite{stuckey}, we restrict ourselves here to the origins of
gauge invariance and QFT.

We understand the reader may not be familiar with the path integral
formalism, as Healey puts it\cite{healey}, ``While many contemporary
physics texts present the path-integral quantization of gauge field
theories, and the mathematics of this technique have been
intensively studied, I know of no sustained critical discussions of
its conceptual foundations.'' Therefore, we begin in section
\ref{section2} with an overview and interpretation of the path
integral formalism, which is particularly well-suited for the study
of gauge invariance.

\section{The Discrete Path Integral Formalism}
\label{section2}

In this section we provide an overview and interpretation of the
path integral approach, showing explicitly how we intend to use
``its conceptual foundations.'' We employ the discrete path integral
formalism because it embodies a 4Dism that allows us to model
spacetimematter. For example, the path integral approach is based on
the fact that\cite{feynman} ``the [S]ource will emit and the
detector receive,'' i.e., the path integral formalism deals with
Sources and sinks as a unity while invoking a description of the
experimental process from initiation to termination. By assuming the
discrete path integral is fundamental to the (conventional)
continuum path integral, we have a graphical basis for the
co-construction of time, space and quantum sources via a
self-consistency criterion (SCC). We will show in section 3 how the
graphical amalgam of spacetimematter underlies QFT.

\subsection{Path Integral in Quantum Physics}

 In the conventional path integral formalism\cite{zee2} for non-relativistic quantum mechanics (NRQM)
 one starts with the amplitude for the propagation from the initial point in configuration space $q_I$ to
 the final point in configuration space $q_F$ in time $T$ via the unitary operator $e^{-iHT}$,
 i.e., $\displaystyle \left \langle q_F \left | e^{-iHT} \right | q_I \right \rangle$. Breaking the
 time $T$ into $N$ pieces $\delta t$ and inserting the identity between each pair of operators $e^{-iH\delta t}$ via
 the complete set $\int dq | q \rangle \langle q | =1$  we have
 \begin{equation*}
 \fl
 \left \langle q_F \left | e^{-iHT} \right | q_I \right \rangle = \left [ \prod_{j=1}^{N-1} \int dq_j \right ]
 \left \langle q_F \left | e^{-iH\delta t} \right | q_{N-1} \right \rangle
 \left \langle q_{N-1} \left | e^{-iH\delta t} \right | q_{N-2} \right \rangle
 \ldots
 \end{equation*}
 \begin{equation*}
 \left \langle q_2 \left | e^{-iH\delta t} \right | q_1 \right \rangle
 \left \langle q_1 \left | e^{-iH\delta t} \right | q_I \right \rangle.
 \end{equation*}
With $H=\hat{p}^2/2m + V(\hat{q})$ and $\delta t \rightarrow 0$ one
can then show that the amplitude is given by
\begin{equation}
\left \langle q_F \left | e^{-iHT} \right | q_I \right \rangle =
\int Dq(t) \exp \left [ i \int_0^T dt L(\dot{q},q) \right ],
\label{eqn1}
\end{equation}
where $L(\dot{q},q) = m \dot{q}^2/2-V(q)$ . If $q$ is the spatial
coordinate on a detector transverse to the line joining Source and
detector, then $\displaystyle \prod_{j=1}^{N-1}$ can be thought of
as $N-1$ ``intermediate'' detector surfaces interposed between the
Source and the final (real) detector, and $\int dq_j$ can be thought
of all possible detection sites on the $j^{\mbox{th}}$ intermediate
detector surface. In the continuum limit, these become $\int Dq(t)$
which is therefore viewed as a ``sum over all possible paths'' from
the Source to a particular point on the (real) detector, thus the
term ``path integral formalism'' for conventional NRQM is often
understood as a sum over ``all paths through space.''

To obtain the path integral approach to QFT one associates $q$ with
the oscillator displacement at a {\em particular point} in space
($V(q) = kq^2/2$). In QFT, one takes the limit $\delta x \rightarrow
0$ so that space is filled with oscillators and the resulting
spatial continuity is accounted for mathematically via $q_i(t)
\rightarrow q(t,x)$, which is denoted $\phi(t,x)$ and called a
``field.'' The QFT transition amplitude $Z$ then looks like
\begin{equation}
Z = \int D\phi \exp \left [ i \int d^4 x L( \dot{\phi}, \phi ) \right ]
\label{eqn2}
\end{equation}
where $L(\dot{\phi},\phi) = (d\phi)^2/2 - V(\phi)$ . Impulses $J$
are located in the field to account for particle creation and
annihilation; these $J$ are called ``sources'' in QFT and we have
$L(\dot{\phi},\phi) = (d\phi)^2/2 - V(\phi) + J(t,x) \phi(t,x)$,
which can be rewritten as $L(\dot{\phi},\phi) = \phi D \phi/2 +
J(t,x) \phi(t,x)$, where $D$ is a differential operator. In its
discrete form (typically, but not necessarily, a hypercubic
spacetime lattice), $D \rightarrow \KKK$ (a difference matrix),
$J(t,x)\rightarrow \JJJ$ (each component of which is associated with
a point on the spacetime lattice) and $\phi \rightarrow \QQQ$ (each
component of which is associated with a point on the spacetime
lattice). Again, part of $\JJJ$ represents field disturbances
emanating from a source location (Source) and the other part
represents field disturbances incident on a source location (sink)
in the conventional view of path integral QFT and, in particle
physics, these field disturbances are the particles. We will keep
the partition of $\JJJ$ into Sources and sinks in our theory X, but
there will be no vacuum lattice structure between the discrete set
of sources. The discrete counterpart to (\ref{eqn2}) is
then\cite{zee3}
\begin{equation}
Z = \int \ldots \int dQ_1 \ldots dQ_N \exp \left[ \frac {i}{2} \QQQ
\cdot \KKK \cdot\QQQ + i \JJJ \cdot\QQQ \right ]. \label{eqn3}
\end{equation}
In conventional quantum physics, NRQM is understood as $(0+1)-$dimensional QFT.

\subsection{Our Interpretation of the Path Integral in Quantum Physics}

 We agree that NRQM is to be understood as $(0+1)-$dimensional QFT, but point out
 this is at conceptual odds with our derivation of (\ref{eqn1}) when $\int Dq(t)$
 represented a sum over all paths in space, i.e., when $q$ was understood as a
 location in space (specifically, a location along a detector surface). If NRQM
 is $(0+1)-$dimensional QFT, then $q$ is a field displacement at a single location
 in space. In that case, $\int Dq(t)$ must represent a sum over all field values at a particular
 point on the detector, not a sum over all paths through space from the Source to a particular
 point on the detector (sink). So, how {\em do} we relate a point on the detector (sink) to the Source?

In answering this question, we now explain a formal difference
between conventional path integral NRQM and our proposed approach:
our links only connect and construct discrete sources $\JJJ$, there
are no source-to-spacetime links (there is no vacuum lattice
structure, only spacetimematter). Instead of $\delta x \rightarrow
0$, as in QFT, we assume $\delta x$ is measureable for (such) NRQM
phenomenon. More specifically, we propose starting with (\ref{eqn3})
whence (roughly) NRQM obtains in the limit $\delta t \rightarrow 0$,
as in deriving (\ref{eqn1}), and QFT obtains in the additional limit
$\delta x \rightarrow 0$, as in deriving (\ref{eqn2}). The QFT limit
is well understood as it is the basis for lattice gauge theory and
regularization techniques, so one might argue that we are simply
{\em clarifying} the NRQM limit where the path integral formalism is
not widely employed. However, again, we are proposing a discrete
starting point for theory X, as in (\ref{eqn3}). Of course, that
discrete spacetime is fundamental while ``the usual continuum theory
is very likely only an approximation\cite{Feinberg}'' is not new.

\subsection{Discrete Path Integral is Fundamental}

The version of theory X we propose is a discrete path integral over
graphs, so  (\ref{eqn3}) {is not a discrete approximation of
(\ref{eqn1}) \& (\ref{eqn2})}, but rather {\em (\ref{eqn1}) \&
(\ref{eqn2}) are continuous approximations of  (\ref{eqn3})}. In the
arena of quantum gravity it is not unusual to find discrete
theories\cite{loll} that are in some way underneath spacetime theory
and theories of ``matter'' such as QFT, e.g., causal dynamical
triangulations\cite{ambjorn}, quantum graphity\cite{konopka} and
causets\cite{sorkin1}. While these approaches are interesting and
promising, the approach taken here for theory X will look more like
Regge calculus quantum gravity (see Bahr \& Dittrich \cite{bahr} and
references therein for recent work along these lines) modified to
contain no vacuum lattice structure.

Placing a discrete path integral at bottom introduces conceptual and
analytical deviations from the conventional, continuum path integral
approach. Conceptually, (\ref{eqn1}) of NRQM represents a sum
over all field values at a particular point on the detector, while
(\ref{eqn3}) of theory X is a mathematical machine that measures
the ``symmetry'' (strength of stationary points) contained in the
core of the discrete action
\begin{equation}
\frac 12 \KKK + \JJJ
\label{eqn4}
\end{equation}
This core or {\em actional} yields the discrete action after
operating on a particular vector $\QQQ$ (field). The actional
represents a {\em fundamental/topological, 4D description of the
experiment} and $Z$ is a measure of its symmetry. [In its Euclidean
form, which is the form we will use, $Z$ is a partition function.]
For this reason we prefer to call $Z$ the symmetry amplitude of the
4D experimental configuration. Analytically, because we are {\em
starting} with a discrete formalism, we are in position to
mathematically explicate trans-temporal identity, whereas this
process is unarticulated elsewhere in physics. As we will now see,
this leads to our proposed self-consistency criterion (SCC)
underlying $Z$.

\subsection{Self-Consistency Criterion} Our use of a self-consistency criterion
is not without precedent, as we already have an ideal example in
Einstein's equations of general relativity (GR). Momentum, force and
energy all depend on spatiotemporal measurements (tacit or
explicit), so the stress-energy tensor cannot be constructed without
tacit or explicit knowledge of the spacetime metric (technically,
the stress-energy tensor can be written as the functional derivative
of the matter-energy Lagrangian with respect to the metric). But, if
one wants a ``dynamic spacetime'' in the parlance of GR, the
spacetime metric must depend on the matter-energy distribution in
spacetime. GR solves this dilemma by demanding the stress-energy
tensor be ``consistent'' with the spacetime metric per Einstein's
equations. For example, concerning the stress-energy tensor, Hamber
and Williams write\cite{Hamber}, ``In general its covariant
divergence is not zero, but consistency of the Einstein field
equations demands $\nabla^{\alpha} T_{\alpha \beta} = 0$ .'' This
self-consistency hinges on divergence-free sources, which finds a
mathematical underpinning in $\partial \partial  = 0$. So,
Einstein's equations of GR are a mathematical articulation of the
boundary of a boundary principle at the classical level, i.e., they
constitute a self-consistency criterion at the classical level, as
are quantum and classical electromagnetism\cite{misner2}. We will
provide an explanation for this fact in section \ref{section3}, but
essentially the graphical SCC of our theory X gives rise to
continuum counterparts in QFT and classical field theory.

In order to illustrate the discrete mathematical co-constuction of
space, time and sources $\JJJ$, we will use graph theory a la
Wise\cite{wise2} and find that $\partial_1\cdot \partial_1^T$, where
$\partial_1$ is a boundary operator in the spacetime chain complex
of our graph satisfying $\partial_1\cdot \partial_2 = 0$ , has
precisely the same form as the difference matrix in the discrete
action for coupled harmonic oscillators. Therefore, we are led to
speculate that $\KKK \propto
\partial_1\cdot \partial_1^T$. Defining the source vector  $\JJJ$
relationally via $\JJJ \propto \partial_1\cdot \eee$ then gives
tautologically per $\partial_1\cdot \partial_2 = 0$ both a
divergence-free $\JJJ$ and $\KKK\cdot \vvv \propto \JJJ$, where
$\eee$ is the vector of links and $\vvv$ is the vector of vertices.
$\KKK\cdot \vvv \propto \JJJ$ is our SCC following from
$\partial_1\cdot\partial_2 = 0$, and it defines what is meant by a
self-consistent co-construction of space, time and divergence-free
sources $\JJJ$, thereby constraining $\KKK$ and $\JJJ$ in $Z$. Thus,
our SCC provides a basis for the discrete action and supports our
view that (\ref{eqn3}) is fundamental to (\ref{eqn1}) \&
(\ref{eqn2}), rather than the converse. Conceptually, that is the
basis of our discrete, graphical path integral approach to theory X.
We now provide the details.

\section{The Formalism}
\label{section3}

\subsection{The General Approach}

\begin{figure}
\begin{center}
\includegraphics[height=30mm]{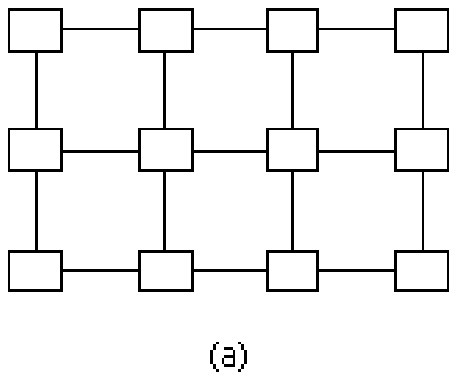} \hspace{10mm}
\includegraphics[height=60mm]{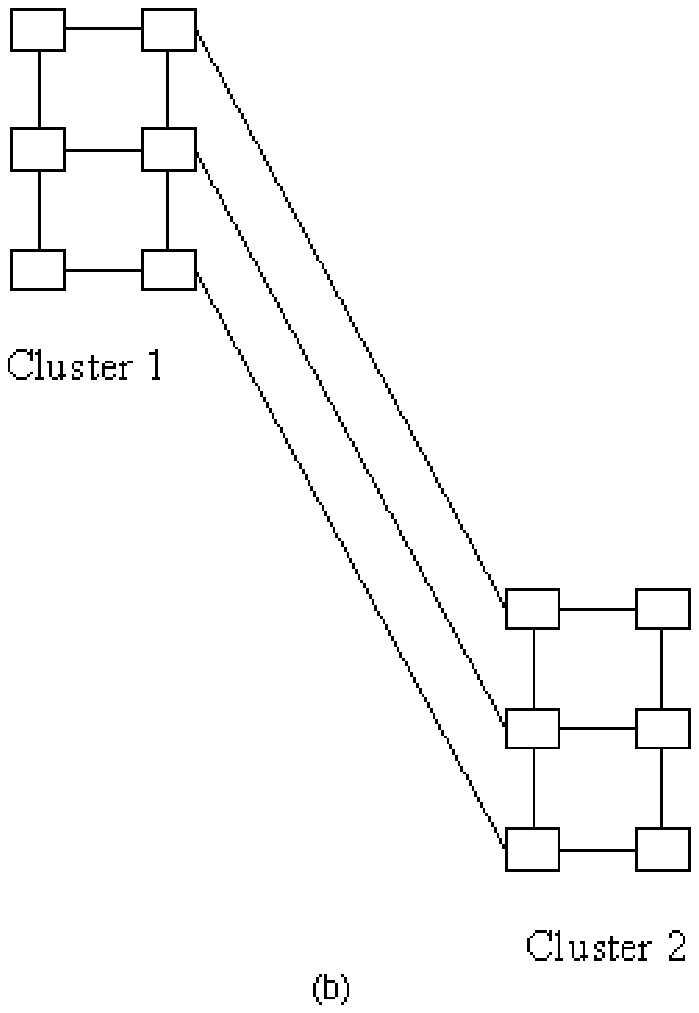}
\end{center}
\caption{(a) Topological Graph - This spacetimematter graph depicts
four sources, i.e., the columns of squares. The graph's actional
$\KKK/2+\JJJ$, such that $\KKK\cdot \vvv \propto \JJJ$,
characterizes the graphical topology, which underwrites a partition
function $Z$ for spatiotemporal geometries over the graph. (b)
Geometric Graph - The topological graph of (a) is endowed with a
particular distribution of spatiotemporal geometric relations, i.e.,
link lengths as determined by the field values $Q$ on their
respective vertices. Clusters 1 \& 2 are the result of this
geometric process for a particular distribution of field values
$Q$.} \label{fig1}
\end{figure}

\begin{figure}
\begin{center}
\includegraphics[height=50mm]{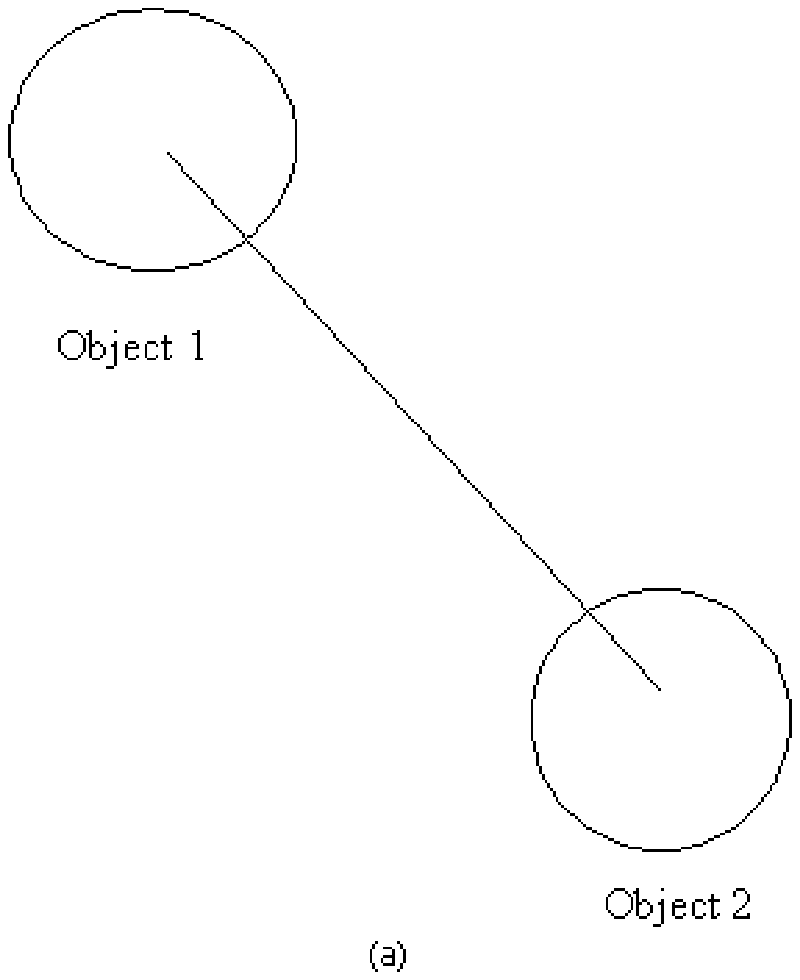} \hspace{10mm}
\includegraphics[height=50mm]{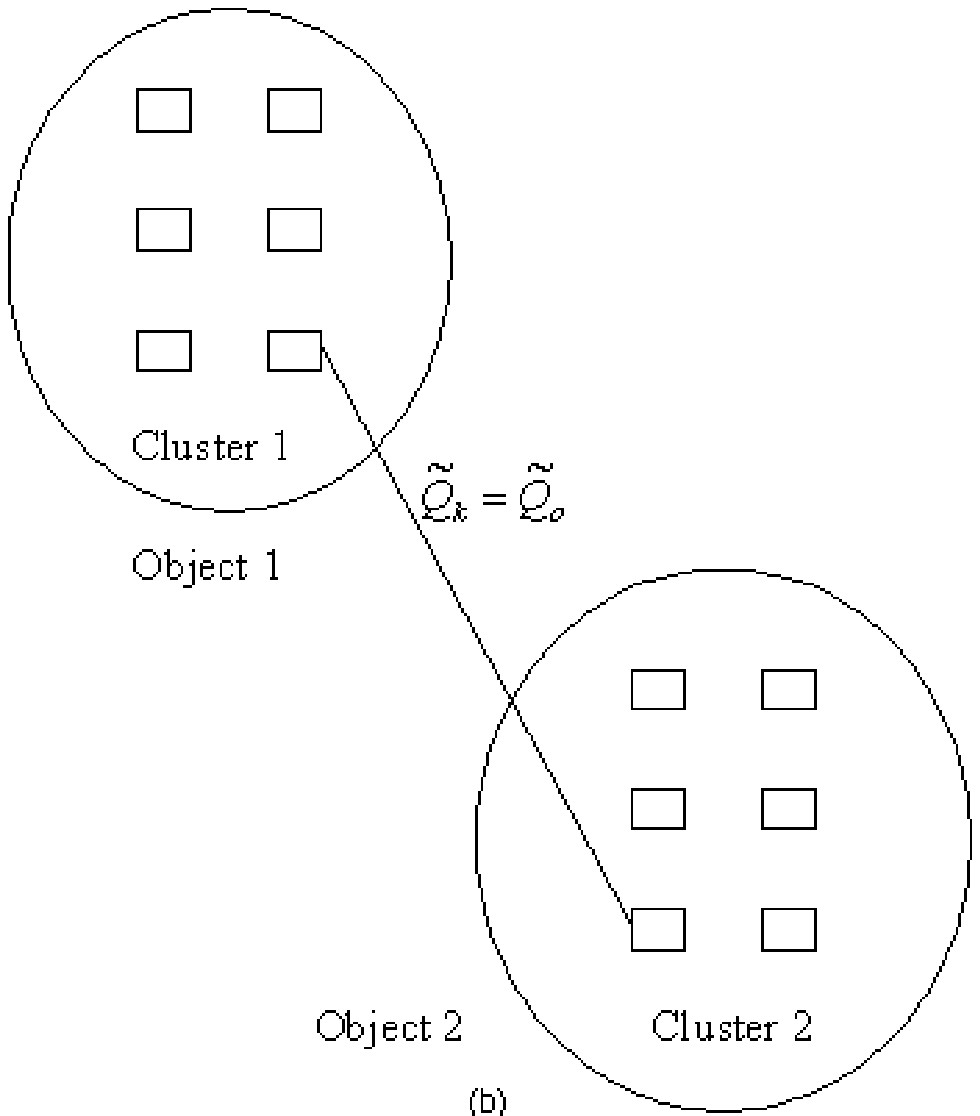}
\end{center}
\caption{(a) Classical Physics - Classical Objects result when the
most probable field values $\QQQ_0$ yield spatiotemporally localized
Clusters 1 \& 2 as in Figure \ref{fig1}b. The lone link in this
figure represents the average of the link lengths obtained via the
most probable field values $\QQQ_0$. The most probable values
$\QQQ_0$ are found via $\KKK \cdot \QQQ_0 = \JJJ$, so this is the
origin of classical physics. (b) Quantum Physics - A particular
outcome $\QQ_0$ of a quantum physics experiment allows one to
compute the $k^{\mbox{th}}$ link length of the geometric graph in
the context of the classical Objects comprising the experiment,
e.g., Source, beam splitters, mirrors, and detectors. The partition
function provides the probability of this particular outcome, i.e.,
$\displaystyle P(\QQ_k = \QQ_0) = \frac {Z(\QQ_k = \QQ_0)}{Z}$.}
\label{fig2}
\end{figure}

Again, in theory X, the symmetry amplitude $Z$ contains a discrete
action constructed per a self-consistency criterion (SCC) for space,
time and divergence-free sources $\JJJ$. As introduced in section
\ref{section2} and argued later in this section, we will codify the
SCC using $\KKK$ and $\JJJ$; these elements are germane to the
transition amplitude $Z$ in the Central Identity of Quantum Field
Theory\cite{zee4},
\begin{equation}
\fl
Z = \int D \ppphi \exp \left [ - \frac 12 \ppphi \cdot \KKK \cdot \ppphi - V(\ppphi) + \JJJ \cdot \ppphi \right ] \\
= \exp \left [ -V \left ( \frac {\delta}{\delta J} \right ) \right ] \exp \left [\frac 12 \JJJ \cdot \KKK^{-1} \cdot \JJJ \right ].
\label{eqn5}
\end{equation}
While the field is a mere integration variable used to produce $Z$,
it must reappear at the level of classical field theory. To see how
the field makes it appearance per theory X, consider (\ref{eqn5}) for the simple Gaussian theory ($V(\phi) = 0$). On a
graph with $N$ vertices, (\ref{eqn5}) is
\begin{equation}
Z = \int_{-\infty}^{\infty} \ldots \int_{-\infty}^{\infty} dQ_1
\ldots dQ_N  \exp \left [-\frac 12 \QQQ \cdot \KKK \cdot \QQQ + \JJJ
\cdot \QQQ \right ] \label{eqn6}
\end{equation}
with a solution of
\begin{equation}
Z = \left ( \frac {(2\pi)^N}{\det \KKK} \right )^{1/2} \exp \left [\frac 12 \JJJ \cdot \KKK^{-1} \cdot \JJJ \right ].
\label{eqn7}
\end{equation}
It is easiest to work in an eigenbasis of $\KKK$ and (as will argue
later) we restrict the path integral to the row space of $\KKK$,
this gives
\begin{equation}
Z = \int_{-\infty}^{\infty} \ldots \int_{-\infty}^{\infty} d\QQ_1
\ldots d\QQ_{N-1}  \exp \left [\sum_{j=1}^{N-1} \left (-\frac 12
\QQ_j^2 a_j + \JJ_j \QQ_j \right ) \right ] \label{eqn8}
\end{equation}
where $\QQ_j$ are the coordinates associated with the eigenbasis of
$\KKK$ and $\QQ_N$ is associated with eigenvalue zero, $a_j$ is the
eigenvalue of $\KKK$  corresponding to $\QQ_j$, and $\JJ_j$ are the
components of $\JJJ$ in the eigenbasis of $\KKK$. The solution of
(8) is
\begin{equation}
Z = \left ( \frac {(2\pi)^{N-1}}{\prod_{j=1}^{N-1} a_j} \right )^{1/2} \prod_{j=1}^{N-1} \exp \left ( \frac {\JJ_j^2}{2a_j} \right ).
\label{eqn9}
\end{equation}
On our view, the experiment is described fundamentally by $\KKK$ and
$\JJJ$ on our topological graph. Again, per (\ref{eqn9}), there is
no field $\QQ$ appearing in $Z$ at this level, i.e., $\QQ$ is only
an integration variable.  $\QQ$ makes its first appearance as
something more than an integration variable when we produce
probabilities from $Z$. That is, since we are working with a
Euclidean path integral, $Z$ is a partition function and the
probability of measuring $\QQ_k=\QQ_0$ is found by computing the
fraction of $Z$ which contains $\QQ_0$ at the $k^{\mbox{th}}$
vertex\cite{lisi}. We have
\begin{equation}
\fl
P \left ( \QQ_k = \QQ_0 \right ) = \frac {Z \left ( \QQ_k = \QQ_0 \right )}{Z} = \sqrt{\frac {a_k}{2\pi}} \exp \left ( - \frac 12 \QQ_0^2 a_k + \JJ_k \QQ_0 - \frac {\JJ_k^2}{2a_k} \right )
\label{eqn10}
\end{equation}
as the part of theory X approximated in the continuum by QFT. The
most probable value of $\QQ_0$ at the $k^{\mbox{th}}$ vertex is then
given by
\begin{equation}
\fl
\delta P \left ( \QQ_k = \QQ_0 \right ) = 0 \Longrightarrow  \delta \left ( - \frac 12 \QQ_0^2 a_k + \JJ_k \QQ_0 - \frac {\JJ_k^2}{2a_k} \right ) = 0 \Longrightarrow a_k \QQ_0 = \JJ_k.
\label{eqn11}
\end{equation}
That is, $\KKK \cdot \QQQ_0 = \JJJ$ is the part of theory X that
obtains statistically and is approximated in the continuum by
classical field theory. We note that the manner by which $\KKK \cdot
\QQQ_0 = \JJJ$ follows from $P(\QQ_k = \QQ_0) = Z(\QQ_k = \QQ_0)/Z$
parallels the manner by which classical field theory follows from
QFT via the stationary phase method\cite{zee5}. Thus, one may obtain
classical field theory by the continuum limit of $\KKK \cdot \QQQ_0
= \JJJ$ in theory X (theory X $\rightarrow$ classical field theory),
or by first obtaining QFT via the continuum limit of $P(\QQ_k =
\QQ_0) = Z(\QQ_k = \QQ_0)/Z$ in theory X and then by using the
stationary phase method on QFT (theory X $\rightarrow$ QFT
$\rightarrow$ classical field theory). In either case, QFT is not
quantized classical field theory in our approach. In summary:

\begin{enumerate}
\item $Z$ is a partition function for an experiment described topologically by $\KKK/2+ \JJJ$  (Figure \ref{fig1}a).
\item $P(\QQ_k = \QQ_0) = Z(\QQ_k = \QQ_0)/Z$ gives us the probability for a particular geometric outcome in that experiment (Figures \ref{fig1}b and \ref{fig2}b).
\item $\KKK\cdot \QQQ_0 = \JJJ$ gives us the most probable values of the experimental outcomes which are then averaged to produce the geometry for the experimental procedure at the classical level (Figure \ref{fig2}a).
\item $P(\QQ_k = \QQ_0) = Z(\QQ_k = \QQ_0)/Z$ and $\KKK\cdot \QQQ_0 = \JJJ$ are the parts of theory X approximated in the continuum by QFT and classical field theory, respectively.
\end{enumerate}

\subsection{The Two-Source Euclidean Symmetry Amplitude/Partition Function}

\begin{figure}
\begin{center}
\includegraphics[height=50mm]{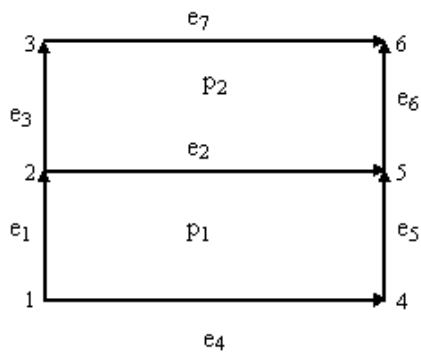}
\end{center}
\caption{Graph with six vertices, seven links $e_i$ and two
plaquettes $p_i$.} \label{fig3}
\end{figure}

Typically, one identifies fundamentally interesting physics with
symmetries of the action in the Central Identity of Quantum Field
Theory, but we have theory X fundamental to QFT, so our method of
choosing fundamentally interesting physics must reside in the
topological graph of theory X. Thus, we seek a constraint of $\KKK$
and $\JJJ$ in our graphical symmetry amplitude $Z$ and this will be
in the form of a self-consistency criterion (SCC). In order to
motivate our general method, we will first consider a simple graph
with six vertices, seven links and two plaquettes for our
$(1+1)-$dimensional spacetime model (Figure \ref{fig3}). Our goal
with this simple model is to seek relevant structure that might be
used to infer an SCC. We begin by constructing the boundary
operators over our graph.

The boundary of $\bfp_1$ is $\bfe_4 + \bfe_5 - \bfe_2 - \bfe_1$,
which also provides an orientation. The boundary of $\bfe_1$ is
$\bfv_2 - \bfv_1$, which likewise provides an orientation. Using
these conventions for the orientations of links and plaquettes we
have the following boundary operator for $C_2 \rightarrow C_1$,
i.e., space of plaquettes mapped to space of links in the spacetime
chain complex:
\begin{equation}
\partial_2 = \left [ \begin{array}{rr}
-1 & 0 \\
-1 & 1 \\
 0 & -1 \\
 1 & 0 \\
 1 & 0 \\
 0 & 1 \\
 0 & -1 \end{array} \right ]
\label{eqn12}
\end{equation}
Notice the first column is simply the links for the boundary of
$\bfp_1$ and the second column is simply the links for the boundary
of $\bfp_2$. We have the following boundary operator for $C_1
\rightarrow C_0$, i.e., space of links mapped to space of vertices
in the spacetime chain complex:
\begin{equation}
\partial_1 = \left [ \begin{array}{rrrrrrr}
-1 & 0 & 0 & -1 & 0 & 0 & 0 \\
 1 & -1 & -1 & 0 & 0 & 0 & 0 \\
 0 & 0 & 1 & 0 & 0 & 0 & -1 \\
 0 & 0 & 0 & 1 & -1 & 0 & 0 \\
 0 & 1 & 0 & 0 & 1 & -1 & 0 \\
 0 & 0 & 0 & 0 & 0 & 1 & 1 \end{array} \right ]
\label{eqn13}
\end{equation}
which completes the spacetime chain complex, $C_0 \leftarrow C_1
\leftarrow C_2$. Notice the columns are simply the vertices for the
boundaries of the edges. These boundary operators satisfy
$\partial_1\cdot\partial_2 = 0$, i.e., the boundary of a boundary
principle.

\begin{figure}
\begin{center}
\includegraphics[height=30mm]{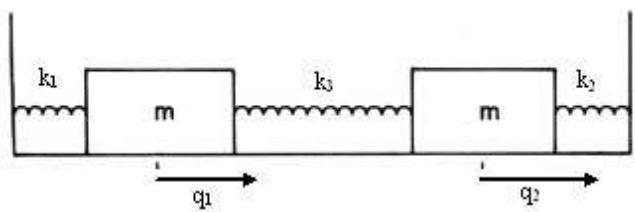}
\end{center}
\caption{Coupled harmonic oscillators.} \label{fig4}
\end{figure}

The potential for coupled oscillators can be written
\begin{equation}
V(q_1,q_2) = \sum_{a,b} \frac 12 k_{ab} q_a q_b = \frac 12 k q_1^2 + \frac 12 k q_2^2 + k_{12} q_1 q_2
\label{eqn14}
\end{equation}
where $k_{11} = k_{22} = k>0$ and $k_{12} = k_{21}<0$ per the
classical analogue (Figure \ref{fig4}) with $k = k_1 + k_3 = k_2 +
k_3$ and $k_{12} = -k_3$ to recover the form in (\ref{eqn14}).
The Lagrangian is then
\begin{equation}
L = \frac 12 m \dot{q}_1^2 + \frac 12 m \dot{q}_2^2 - \frac 12
kq_1^2 - \frac 12 k q_2^2 - k_{12} q_1q_2 \label{eqn15}
\end{equation}
so our NRQM Euclidean symmetry amplitude is
\begin{equation}
\fl
Z = \int Dq(t) \exp \left [ - \int_0^T dt \left ( \frac 12 m \dot{q}_1^2 + \frac 12 m \dot{q}_2^2 + V(q_1, q_2) - J_1 q_1 - J_2 q_2 \right )\right ]
\label{eqn16}
\end{equation}
after Wick rotation. This gives
\begin{equation}
\fl \KKK = \left [ \begin{array}{rrrrrr}
\left ( \frac m{\Delta t} + k \Delta t \right ) & -\frac m{\Delta t} & 0 & k_{12} \Delta t & 0 & 0 \\
-\frac m{\Delta t} & \left ( \frac {2m}{\Delta t} + k \Delta t \right ) & -\frac m{\Delta t} & 0 & k_{12} \Delta t & 0 \\
0 & -\frac m{\Delta t} & \left ( \frac m{\Delta t} + k \Delta t \right ) & 0 & 0 & k_{12} \Delta t \\
k_{12} \Delta t & 0 & 0 & \left ( \frac m{\Delta t} + k \Delta t \right ) & -\frac m{\Delta t} & 0 \\
0 & k_{12} \Delta t & 0 & -\frac m{\Delta t} & \left ( \frac {2m}{\Delta t}  + k \Delta t \right ) & -\frac m{\Delta t} \\
0 & 0 & k_{12} \Delta t & 0 & -\frac m{\Delta t} & \left ( \frac m{\Delta t} + k \Delta t \right ) \end{array} \right ]
\label{eqn17}
\end{equation}
on our graph. Thus, we borrow (loosely) from Wise\cite{wise3} and
suggest $\KKK \propto \partial_1\cdot\partial_1^T$ since
\begin{equation}
\partial_1\cdot\partial_1^T = \left [ \begin{array}{rrrrrr}
2 & -1 & 0 & -1 & 0 & 0 \\
-1 & 3 & -1 & 0 & -1 & 0 \\
0 & - 1& 2 & 0 & 0 & -1 \\
-1 & 0 & 0 & 2 & -1 & 0 \\
0 & -1 & 0 & -1 & 3 & -1 \\
0 & 0 & -1 & 0 & -1 & 2 \end{array} \right ]
\label{eqn18}
\end{equation}
produces precisely the same form as (\ref{eqn17}) and quantum
theory is known to be ``rooted in this harmonic
paradigm\cite{zee6}.'' [In fact, these matrices will continue to
have the same form as one increases the number of vertices in Figure
\ref{fig3}.] Now we construct a suitable candidate for $\JJJ$,
relate it to $\KKK$ and infer our SCC.

Recall that $\JJJ$  has a component associated with each vertex so
here it has components, $J_n$, $n = 1, 2, \ldots, 6$; $J_n$ for $n =
1, 2, 3$ represents one source and $J_n$ for $n = 4, 5, 6$
represents the second source. We propose $\JJJ \propto
\partial_1\cdot\eee$, where $e_i$ are the links of our graph, since
\begin{equation}
\fl
\partial_1\cdot\eee =
\left [ \begin{array}{rrrrrrr}
-1 & 0 & 0 & -1 & 0 & 0 & 0 \\
 1 & -1 & -1 & 0 & 0 & 0 & 0 \\
 0 & 0 & 1 & 0 & 0 & 0 & -1 \\
 0 & 0 & 0 & 1 & -1 & 0 & 0 \\
 0 & 1 & 0 & 0 & 1 & -1 & 0 \\
 0 & 0 & 0 & 0 & 0 & 1 & 1 \end{array} \right ]
\left [ \begin{array}{c} e_1 \\ e_2 \\ e_3 \\ e_4 \\ e_5 \\ e_6 \\ e_7 \end{array} \right ]
 = \left [\begin{array}{c} -e_1-e_4 \\ e_1 - e_2-e_3 \\ e_3 - e_7 \\ e_4 - e_5 \\ e_2 + e_5 - e_6 \\ e_6 + e_7 \end{array}\right ]
\label{eqn19}
\end{equation}
automatically makes $\JJJ$ divergence-free, i.e., $\displaystyle
\sum_i J_i = 0$. Such a relationship on discrete spacetime lattices
is not new. For example, Sorkin showed that charge conservation
follows from gauge invariance for the electromagnetic field on a
simplicial net\cite{sorkin2}.

With these definitions of $\KKK$  and $\JJJ$  we have, ipso facto,
$\KKK\cdot \vvv \propto \JJJ$  as the basis of our SCC since
\begin{equation}
\fl
\partial_1\cdot\partial_1^T \cdot \vvv = \left [ \begin{array}{rrrrrr}
2 & -1 & 0 & -1 & 0 & 0 \\
-1 & 3 & -1 & 0 & -1 & 0 \\
0 & - 1& 2 & 0 & 0 & -1 \\
-1 & 0 & 0 & 2 & -1 & 0 \\
0 & -1 & 0 & -1 & 3 & -1 \\
0 & 0 & -1 & 0 & -1 & 2 \end{array} \right ] \left [
\begin{array}{c} v_1 \\ v_2 \\ v_3 \\ v_4 \\ v_5 \\ v_6 \end{array}
\right ] = \left [\begin{array}{c} -e_1-e_4 \\ e_1 - e_2-e_3 \\ e_3 - e_7 \\
e_4 - e_5\\ e_2 + e_5 - e_6 \\ e_6 + e_7 \end{array} \right ] =
\partial_1\cdot\eee \label{eqn20}
\end{equation}
where we have used $e_1 = v_2 - v_1$ (etc.) to obtain the last
column. You can see that the boundary of a boundary principle
underwrites (\ref{eqn20}) by the definition of ``boundary'' and
from the fact that the links are directed and connect one vertex to
another, i.e., they do not start or end `off the graph'. Likewise,
this fact and our definition of $\JJJ$ imply $\displaystyle \sum_i
J_i = 0$, which is our graphical equivalent of a divergence-free,
relationally defined source (every link leaving one vertex goes into
another vertex). Thus, the SCC $\KKK\cdot \vvv \propto \JJJ$ and
divergence-free sources $\displaystyle \sum_i J_i = 0$ obtain
tautologically via the boundary of a boundary principle. The SCC
also guarantees that $\JJJ$ resides in the row space of $\KKK$  so,
as will be shown, we can avoid having to ``throw away infinities''
associated with gauge groups of infinite volume as in Faddeev-Popov
gauge fixing. $\KKK$ has at least one eigenvector with zero
eigenvalue which is responsible for gauge invariance, so {\em the
self-consistent co-construction of space, time and divergence-free
sources entails gauge invariance.}

Moving now to $N$ dimensions, the Wick rotated version of (\ref{eqn3}) is (\ref{eqn6}) and the solution is (\ref{eqn7}). Using $\JJJ = \alpha
\partial_1\cdot\eee$  and $\KKK = \beta \partial_1\cdot
\partial_1^T$  ($\alpha, \beta \in \mathbb{R}$) with the SCC gives
$\KKK\cdot \vvv = (\beta/\alpha) \JJJ$, so that $\vvv =
(\beta/\alpha) \KKK^{-1}\cdot \JJJ$. However, $\KKK^{-1}$  does not
exist because $\KKK$ has a nontrivial null space, therefore the row
space of $\KKK$ is an $(N-1)-$dimensional subspace of the
$N-$dimensional vector space\footnote{This assumes the number of
degenerate eigenvalues always equals the dimensionality of the
subspace spanned by their eigenvectors, which we will see is true
for $\KKK$ in this example. }. The eigenvector with eigenvalue of
zero, i.e., normal to this hyperplane, is $\left[
\begin{array}{ccccc} 1 & 1 & 1 & \ldots & 1 \end{array} \right ]^T$,
which follows from the SCC as shown supra. Since $\JJJ$ resides in
the row space of $\KKK$ and, on our view, $Z$ is a functional of
$\KKK$ and $\JJJ$ which produces a partition function for the
various $\KKK/2+\JJJ$  associated with different 4D experimental
configurations, we restrict the path integral of (\ref{eqn6}) to the
row space of $\KKK$. Thus, our approach revises (\ref{eqn7}) to give
(\ref{eqn9}).

We find in general that half the eigenvectors of $\KKK$ are of the
form $\displaystyle \left [ \begin{array}{c} \xxx \\ \xxx \end{array} \right ]$
and half are of the form $\displaystyle \left [ \begin{array}{c} \xxx \\
-\xxx \end{array} \right ]$. The eigenvalues are given by $\lambda \pm 1$
where $\lambda - 1$ is the eigenvalue for $\displaystyle
\left [ \begin{array}{c} \xxx \\ \xxx \end{array} \right ]$, $\lambda + 1$ is the
eigenvalue for $\displaystyle \left [ \begin{array}{c} \xxx \\ -\xxx
\end{array} \right ]$, and $\lambda_j = 3-2\cos (2j\pi/N)$,
$j=0,\ldots,N/2-1$. The $k$ components of $\xxx$ for a given
$\lambda_j$ are $\displaystyle x_{jk} = \sqrt{\frac 2N} \cos \left (
\frac {j(2k-1)\pi}{N} \right )$, $k=1,\ldots,N/2$ for $j>0$ and
$x_{0k} = 1/\sqrt{N}$, $k=1,\ldots,N/2$ for $j = 0$ ($j = 0
\rightarrow$ eigenvalues of $\KKK$ are $0$ and $2$). As you can see,
there are no degeneracies within the $\displaystyle \left [ \begin{array}{c}
\xxx \\ \xxx
\end{array} \right ]$ subspace or the $\displaystyle \left [ \begin{array}{c} \xxx
\\ -\xxx \end{array} \right ]$ subspace. Therefore, the only degeneracies
occur between subspaces, so we know all degenerate eigenvalues are
associated with unique eigenvectors, as alluded to in a previous
footnote.

We have $N$ vertices and $(3N/2 - 2)$ links. Define the temporal
(vertical) links $e_i$ in terms of vertices $v_i$ in the following
fashion: $e_i = v_{i+1}-v_i$, $i=1, \ldots, N/2-1$ and
$\displaystyle e_{N/2+i-1} = v_{N/2+i+1}-v_{N/2+i}$, $i=1, \ldots,
N/2-1$.  Define the spatial (horizontal) links via: $\displaystyle
e_{N + i - 2} = v_{N/2+i} - v_i$, $i=1,\ldots,N/2$. This gives
\begin{equation}
\JJJ = \left [ \begin{array}{cc}
-e_1 - e_{N-1} & \\
-e_i + e_{i-1} - e_{N+i-2} & i=2,\ldots \frac N2 -1 \\
e_{N/2-1} - e_{N+N/2-2} & \\
e_{N-1} - e_{N/2} & \\
e_{N/2+i-2}+e_{N+i-2}-e_{N/2+i-1} & i=2,\ldots,\frac N2 -1 \\
e_{N+N/2-2} + e_{N-2}
\end{array} \right ].
\label{eqn21}
\end{equation}
We then need to find the projection of $\JJJ$ on each of the
orthonormal eigenvectors of $\KKK$ that have non-zero eigenvalues.
Call each projection $\JJ_i = \langle i | J \rangle$, where $\langle
i |$ is the $i^{\mbox{th}}$ orthonormal eigenvector. Let $a_i$ ($i =
1, \ldots, N-1$) be the non-zero eigenvalues of $\KKK$ associated
with the eigenvectors $\langle i |$ , ($i = 1,\ldots, N-1$),
respectively. To complete the two-source Euclidean symmetry
amplitude we need to compute the exponent
\begin{equation}
\Phi = \sum_{i=1}^{N-1} \frac {\left ( \JJ_i \right )^2}{2 a_i \hbar \beta}
\label{eqn22}
\end{equation}
where $\hbar$ is viewed as a fundamental scaling factor with the
dimensions of action. We find $\Phi = (\Phi_S + \Phi_T +
\Phi_{ST})/(2\hbar \beta)$, where
\begin{equation}
\Phi_S = \frac {2 \alpha^2}{N} \left ( \sum_{k=1}^{N/2} e_{k+N-2} \right )^2
\label{eqn23}
\end{equation}
involves only spatial links
\begin{equation}
\Phi_T = \frac {2 \alpha^2}{N} \sum_{j=1}^{N/2-1} \left [ \sum_{k=1}^{N/2-1} \left ( e_k + e_{k+N/2-1} \right )\sin \left ( \frac {2jk\pi}N \right ) \right ]^2
\label{eqn24}
\end{equation}
involves only temporal links and
\begin{equation*}
\fl
\Phi_{ST} = \sum_{j=1}^{N/2-1} \frac {4\alpha^2}{N \left ( 1 + 2 \sin^2 \left (\frac {j\pi}N \right)\right )} \left [ \sin \left (\frac {j \pi}N\right)\sum_{k=1}^{N/2-1} \left ( e_k - e_{k+N/2-1} \right ) \sin \left ( \frac {2jk\pi}N \right ) \right .
\end{equation*}
\begin{equation}
\left.+\sum_{k=1}^{N/2}e_{k+N-2}\cos\left(\frac{(2k-1)j\pi}{N}\right)\right]^2
\label{eqn25}
\end{equation}
involves a mix of spatial and temporal links. (\ref{eqn23})
comes from the eigenvalue $2$ associated with $\displaystyle
\left [ \begin{array}{c} \xxx \\ -\xxx \end{array} \right ]$, which exists for all
$N$ under consideration. (\ref{eqn25}) comes from the remaining
eigenvalues associated with $\displaystyle \left [ \begin{array}{c}  \xxx \\
-\xxx \end{array} \right ]$. (\ref{eqn24}) comes from the eigenvalues
associated with $\displaystyle \left [ \begin{array}{c}  \xxx \\ \xxx
\end{array} \right ]$ having omitted zero, which exists for all $N$ under
consideration.

\section{Conclusion}

We have assumed the existence of a discrete theory (X) fundamental
to quantum physics, the characteristics of which we articulated and
explored via a path integral formalism over graphs. Mathematically,
one can summarize our proposed theory X as follows:
\begin{displaymath}
\fl
\KKK\cdot \vvv \propto \JJJ \rightarrow \frac 12 \KKK + \JJJ
\rightarrow Z \rightarrow P \left ( \QQ_k = \QQ_0 \right ) = \frac
{Z\left ( \QQ_k = \QQ_0 \right )}{Z} \rightarrow \KKK\cdot \QQQ_0 =
\JJJ
\end{displaymath}
with QFT and classical field theory understood as the continuum
approximations to $\displaystyle P \left ( \QQ_k = \QQ_0 \right ) =
\frac {Z\left ( \QQ_k = \QQ_0 \right )}{Z}$  and $\KKK\cdot \QQQ_0 =
\JJJ$, respectively. Thus, the graphical SCC $\KKK\cdot \vvv \propto
\JJJ$  statistically reproduces its counterpart $\KKK\cdot \QQQ_0 =
\JJJ$ whence classical field theory. While the mathematical details
of theory X provided herein are too simplistic to unify physics
formally, we do believe they provide a respectable conceptual
response to Glashow and Rovelli's challenges presented in section 1.
Our proposed new approach to theory X underlying QFT accounts
naturally for gauge invariance via a self-consistency criterion and
deals effectively with factors of infinity associated with gauge
groups of infinite volume, since the transition amplitude Z is
restricted to the row space of the difference matrix and source
vector.

While positing a discrete theory at bottom is hardly unique in
fundamental physics, and our formal development is tentative, our
overall approach to theory X is novel in that it is adynamical and
acausal, in contrast to other fundamental theories such as M-theory,
loop quantum gravity, causets, etc. Such theories may deviate from
the norm by employing radical new fundamental entities (branes,
loops, ordered sets, etc.), but the game is always dynamical,
broadly construed (vibrating branes, geometrodynamics, sequential
growth process, etc.). While itself adynamical, the SCC guarantees
the graph will produce divergence-free classical dynamics in the
appropriate statistical and continuum limits, and provides an
acausal global constraint that results in a self-consistent,
co-construction of space, time and matter that is \emph{de facto}
background independent. Thus in our approach, one has an acausal,
adynamical unity of ``spacetimematter'' at the fundamental level
that results statistically in the causal, dynamical ``spacetime +
matter'' of classical physics. Consequently, fundamental explanation
is in terms of a global, adynamical organizing principle. And,
ultimate explanation in physics is not in terms of some thing or
dynamical entity (obeying a new dynamical equation) ``at the
bottom'' conceived at higher energies and smaller spatiotemporal
scales, begging for justification from something at some yet
``deeper'' scale, but self-consistency writ large for the
explanatory ``process'' as a whole.

\section*{References}

\end{document}